\font\titleftb=cmbx12 at 14.4pt
\font\sectionft=cmr12
\font\sectionftb=cmbx12

\font\smallft=cmr9
\font\smallbft=cmbx9
\nopagenumbers
\hsize=126mm
\vsize=190mm
\hoffset=15mm
\voffset=22.6mm

\vbox to 18pt{}
\centerline{\titleftb Acceleration of UHE Cosmic Ray Particles}
\medskip
\centerline{\titleftb at Relativistic Jets}
\medskip
\centerline{\titleftb in Extragalactic Radio Sources}

\bigskip\bigskip\bigskip
\centerline{\sectionft  M. Ostrowski}
\medskip
\centerline{\it Obserwatorium Astronomiczne, Uniwersytet Jagiello\'nski,}
\centerline{\it Krak\'ow, Poland (E-mail:  mio{@}oa.uj.edu.pl)}

\bigskip
\midinsert
\centerline{\sectionftb Abstract} \par
\medskip
\narrower
\noindent
{\smallft A mechanism of ultra-high energy (UHE) cosmic ray
acceleration in extragalactic radio sources at the interface between
the relativistic jet and the ambient medium is discussed as a
supplement to the shock acceleration in `hot spots'. Particles
accelerated at the jet side boundary are expected to dominate at highest
energies. The spectrum formation near the cut-off energy is modeled
using the Monte Carlo particle simulations.} \endinsert

\bigskip \noindent
{\sectionftb  1. Acceleration processes in relativistic jets}
\medskip

We extend the discussion of particle acceleration at shock waves formed
in the terminal points of relativistic jets (Rachen \& Biermann 1993,
Sigl et al. 1995) by including the additional acceleration process
acting at the jet boundary layer. For particles with UHE energies both
the shock and the transition layer between the jet and the ambient
medium can be approximated as surfaces of discontinuous velocity change.
The tangential discontinuity considered in the latter case can be an
efficient cosmic ray acceleration site if the considered velocity
difference $U$ is relativistic and the sufficient amount of turbulence
is present on its both sides (Berezhko 1990, Ostrowski 1990). On
average, for an individual boundary crossing UHE particle gain energy is

$$<\Delta E> / E \, = \, \rho_e \, (\gamma_u-1)   \qquad ,$$

\noindent
where $\gamma_u$ is the flow Lorentz factor and the numerical factor
$\rho_e$ depends on particle anisotropy at the discontinuity. In the
presence of efficient particle scattering, particle simulations
give $\rho_e$ as a substantial fraction of unity (Ostrowski 1990). Let
us also note that in the case of non-relativistic velocity jump, $U <<
c$, the acceleration process is of the second-order in $U/c$.

\vfill \break

\topinsert
\vskip 9cm  \noindent
{\smallbft Fig.~1}
{\smallft The model applied for the final part of the jet, near the
terminal shock. An example particle trajectory is sketched.} \endinsert

\bigskip
\noindent
{\sectionftb  2. The accelerated particle spectrum}
\medskip

    Spectra of particles accelerated at relativistic shock waves depend
to a large extent on the poorly known physical conditions near the
shock. Therefore, in the present simulations we do not attempt to
reproduce a detailed shape of the particle spectrum, but rather
consider the form of spectrum modifications introduced to the {\it
power-law with a cut-off} shock spectrum by the additional acceleration
at the jet boundary. We neglect the radiation losses, i.e. the upper
energy limit is fixed by the boundary conditions allowing for escape of
highest energy particles.

    In the simulations, we consider the shock resting with respect to
the cocoon surrounding the jet (Fig.~1). The upstream plasma hitting the
shock moves with the relativistic velocity $U_1$ and is advected
downstream with the velocity $U_2$. The compression ratio $R = U_1 /
U_2$ is derived for the shock propagating in cold electron-proton
plasma. The conditions arising behind the jet terminal shock due to
the flow divergence are modeled by imposing a free escape boundary
for particle escape at finite distance, $L_{esc}$, downstream the shock
and in the front part of the cocoon adjoining this boundary. Furthermore, we
introduce another, tube-like free escape boundary surrounding the jet at
a distance $R_{esc}$ from the jet axis. The mean magnetic
field is assumed to be parallel to the jet velocity both within the jet
and in the cocoon. In the examples presented below the ratio of
the {\it effective} cross-field diffusion coefficient to the parallel
diffusion coefficient is $D = 0.13$.

\topinsert
\vskip 8cm  
\par \noindent
{\smallbft Fig.~2}
{\smallft Particle distributions $F(p) \equiv dN(p)/d(log \,  p)$, which
give the particle numbers per logarithmic momentum bandwidth, for
particles escaping through the boundaries. Particles with the momentum
$p = 1.0$ have a gyroradius equal to the jet radius, $R_j$. The
injection momentum is $p_0 = 10^{-3}$, and we take $R_{esc} = 2 R_j$ and
$L_{esc} = 1$ downstream diffusive scale. With shorter dashes we denote
the spectrum for the front boundary, with the longer ones for the side
boundary, and the full line represents a sum of the two. The dashed
straight line is a power-law fit to the low-energy part of the
spectrum.}
\endinsert

\midinsert
\vskip 8cm  
\par \noindent
{\smallbft Fig.~3}
{\smallft A comparison of particle spectra generated in jets with
different velocities $U_1$ given near the respective curves.}
\endinsert

\midinsert
\vskip 8cm  \noindent
{\smallbft Fig.~4}
{\smallft Comparison of the spectra obtained for the far upstream seed
particle injection (a, b) and for the shock injection (c, d). $R_{esc} =
1 R_j$ for dashed lines (b, d) and $10 R_j$ for solid lines (a, c).In
all cases $L_{esc} = 1$.} \endinsert

At Fig-s~2,3 we consider spectra of particles escaping through these
boundaries for the seed particle injection at the shock. The shock
acceleration process determines the spectrum inclination due to the
combined action of the particle energization at the shock and the
continuous particle advection with the plasma. In the present
simulations we consider fixed spatial distances to the escape
boundaries, but the size of particle trajectory defined by its
gyroradius, as well as the spatial diffusion coefficients, increase in
proportion to particle momentum. Due to this increase, the escape
probability grows with the particle momentum providing a cut-off in the
spectrum. The energy scale of the cut-off is different for the shock
spectrum and the side boundary spectrum, with the latter being always
larger. The difference between these two scales increases with e.g. the
jet velocity, the extent of the diffusive cocoon, shifting the particle
injection site upstream the shock, increasing the effective particle
radial diffusion coefficient. One should note that in the range of
particle energies directly preceding the cut-off energy the total
spectrum exhibits some flattening with respect to the inclination
expected for the standard picture of the shock acceleration. There are
two reasons for that flattening: an additional particle transport from
the downstream shock region to the upstream one through the cocoon
surrounding the jet (this effect occurs also if  there is no side
boundary acceleration ! ), and inclusion of the very flat spectral
component resulting from the side boundary acceleration.

At Fig.~4 we consider particles injected far upstream the shock, in the
distance $10^3 \, R_j$, at the jet side boundary. In this case the
resulting distribution of particles is very flat. This feature results
from the character of the acceleration process with particles having
an opportunity to hit the accelerating surface again and again due to
inefficient diffusive escape to the sides. The apparent deficiency of
low-energy particles in the spectrum results from the fact that most of
these particles succeeded in crossing the discontinuity several times before
they were able to diffusively escape through the side boundary. In
another words, the escape due to particle energy increase (and the
corresponding diffusion coefficient increase) is much  more effective than
the escape due to low energy particle diffusion across the cocoon.

The present work was supported by the {\it Komitet Bada\'n Naukowych}
(Project 2~P03D~016~11).

\bigskip \noindent
{\sectionftb  References}
\medskip
\noindent
Berezhko, E.G., 1990, Preprint {\it Frictional Acceleration of Cosmic Rays},
\parskip=0pt \par
The Yakut Scientific Centre, Yakutsk \hfill \break
Ostrowski, M., 1990, A\&A, {\bf 238}, 435 \hfill \break
Rachen, J.P., Biermann, P., 1993, A\&A, {\bf 272}, 161  \hfill \break
Sigl, G., Schramm, D.N., Bhattacharjee, P., 1995. Astropart. Phys. {\bf 2},
\parskip=0pt \par
 401

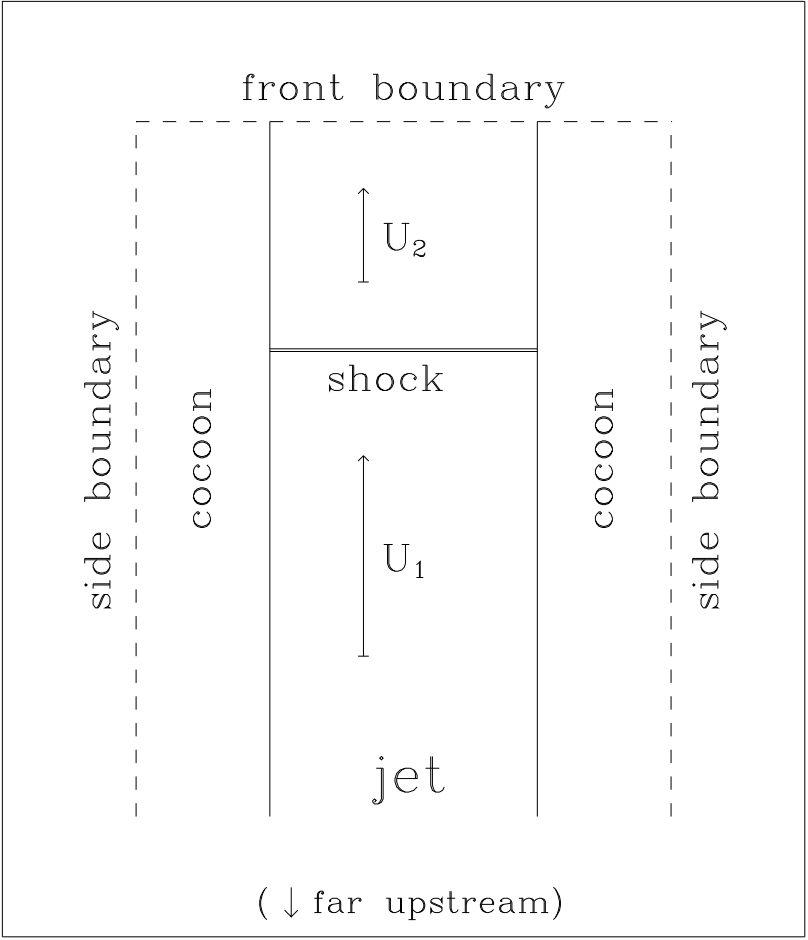
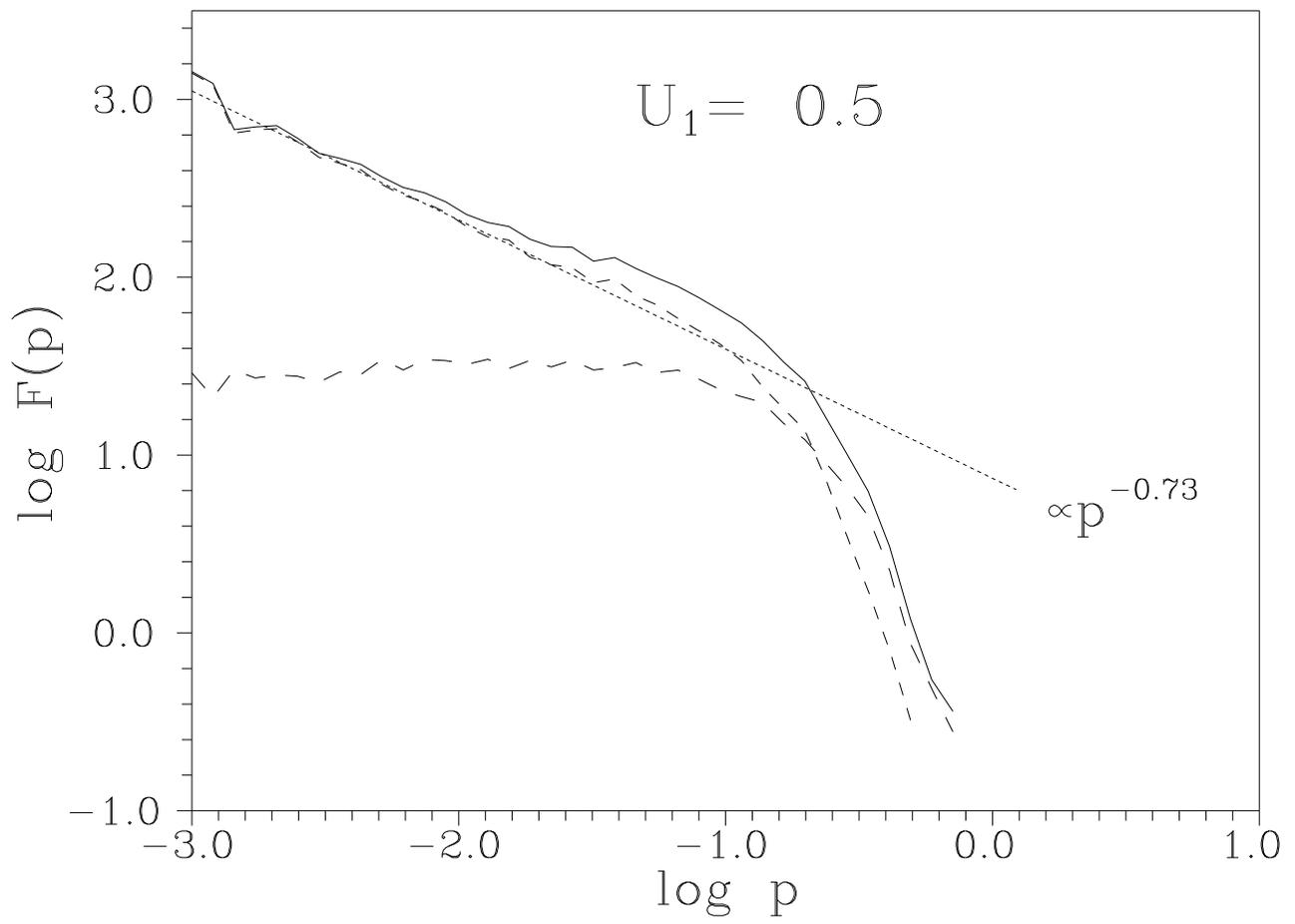
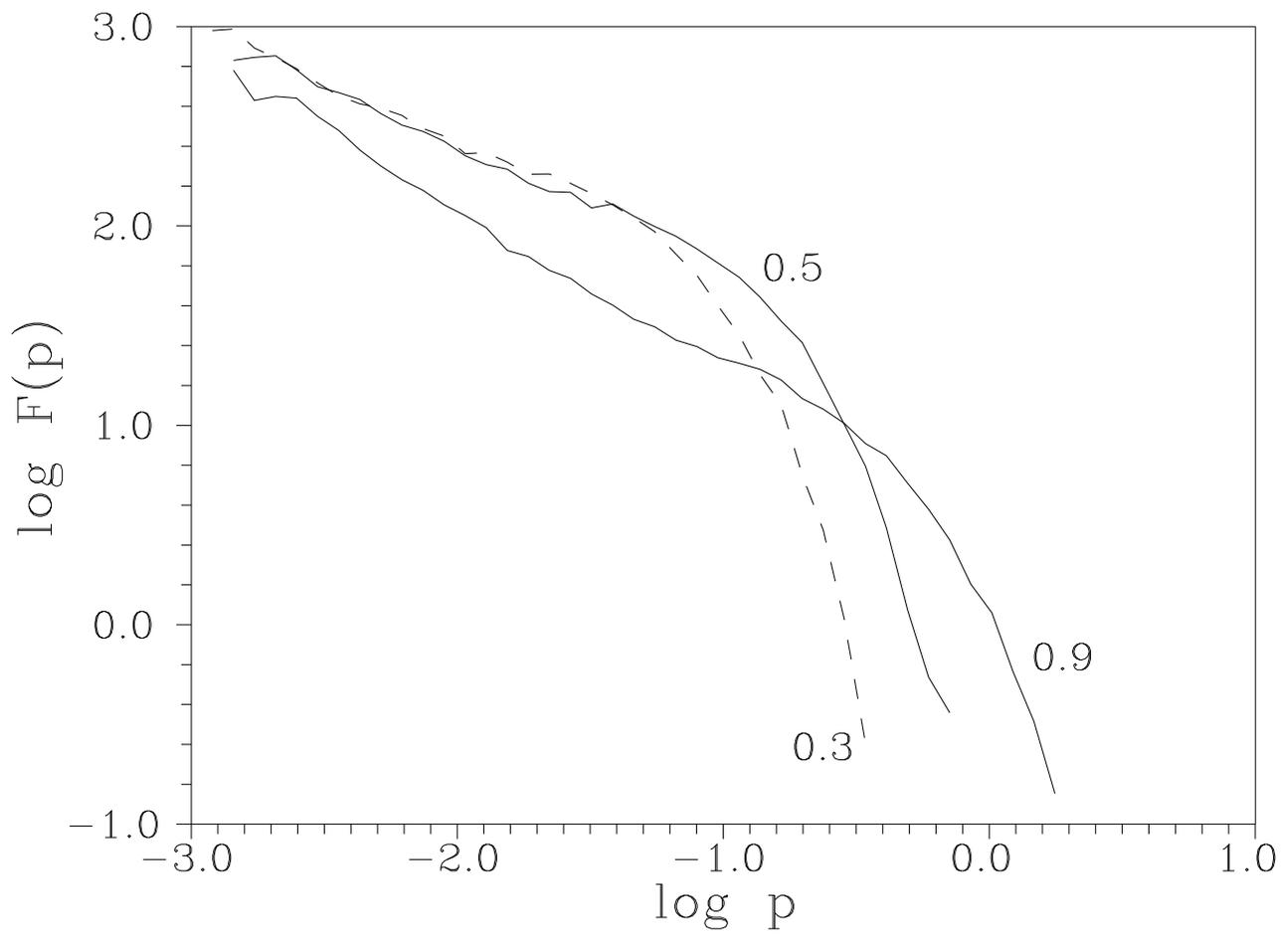
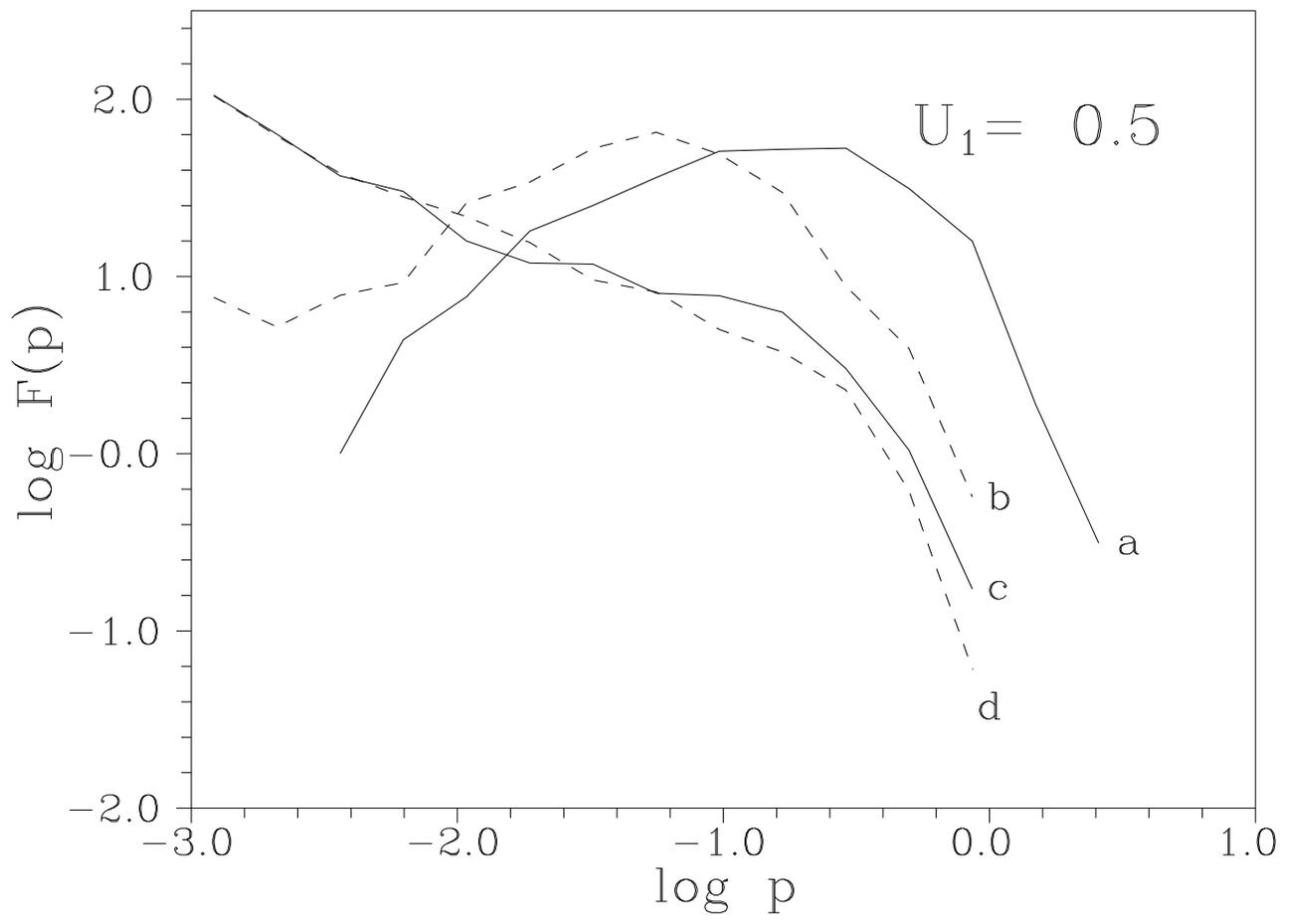
\bye